\documentclass[%
 reprint,
 amsmath,amssymb,
 aps,
 pra,
]{revtex4-1}

\usepackage{graphicx}
\usepackage{dcolumn}
\usepackage{bm}
\usepackage{float}
\usepackage{xcolor}
\usepackage{amsmath}
\usepackage{mathtools}
\usepackage{etoolbox}
\usepackage{physics}
\apptocmd{\thebibliography}{\raggedright}{}{}
\usepackage{amsmath}
\usepackage[titletoc]{appendix}

\usepackage{soul}

\def \beq {\begin{equation}}
\def \eeq {\end{equation}}

\begin{document}


\title {Arbitrary length XX spin chains boundary-driven by non-Markovian environments}

\author{G. Mouloudakis$^{1,2}$}
 \email{gmouloudakis@physics.uoc.gr}

\author{T. Ilias$^{1}$}

\thanks{Current Affiliation: Institut für Theoretische Physik, Albert-Einstein-Allee 11, Universität Ulm, 89069 Ulm, Germany}

\author{P. Lambropoulos$^{1,2}$}%

\affiliation{${^1}$Department of Physics, University of Crete, P.O. Box 2208, GR-71003 Heraklion, Crete, Greece
\\
${^2}$Institute of Electronic Structure and Laser, FORTH, P.O.Box 1527, GR-71110 Heraklion, Greece}

\date{\today}

\begin{abstract}
In this work we provide a recursive method of calculating the wavefunction of a XX spin chain coupled at both ends to  non-Markovian reservoirs with arbitrary spectral density. The method is based on the appropriate handling of the time-dependent Schr\"{o}dinger's equations of motion in Laplace space and leads to closed form solutions of the transformed amplitudes, for arbitrary chain lengths as well as arbitrary initial conditions, within the single-excitation subspace. Results on the dynamical as well as state transfer properties of the system for various combinations of parameters are also presented. In particular, detailed quantitative comparisons for Lorentzian and Ohmic reservoirs are illustrated. 
\end{abstract}

\maketitle

\section{Introduction}

One dimensional many-body systems such as quantum spin chains arise in many contexts throughout quantum information theory as well as condensed matter physics, due to their versatility as basic resources for the implementation of solid-state devices for quantum computing and quantum communication tasks \cite{ref1}. Among these tasks, faithful quantum state transfer \cite{ref2,ref3,ref4,ref5,ref6,ref7,ref8,ref9} and long-distance entanglement \cite{ref10,ref11,ref12,ref13,ref14,ref15} have been investigated for various spin chain configurations in great detail throughout the last 20 years or so, with the research on these fields being still active. 

In recent years, much interest has arisen in the study of properties of open quantum systems interacting with external environments, since dissipation and decoherence  are inevitably present in any experimental realization of such systems. Such systems may consist of just a pair of qubits, with focus on the effects of the environmental dissipation on the generated bipartite entanglement \cite{ref16,ref17,ref18,ref19,ref20}, up to a whole quantum spin chain with focus on state transfer \cite{ref21,ref22,ref23, ref24} or short- and long-distance correlations \cite{ref25,ref26,ref27,ref28}. Extensions that account for the non-Markovian character of the surrounding environment have been also made \cite{ref29,ref30,ref31,ref32,ref33,ref34,ref35}, revealing interesting effects associated with the  memory character  of the reservoir, which enable the exchange of information between the system and the environment within finite times.

At the same time, attention has been drawn to a special class of open quantum systems, i.e. the so-called boundary-driven open quantum systems \cite{ref36}. These are quantum systems coupled  to external environments at their edges and are usually studied for their transport \cite{ref37,ref38,ref39,ref40} and thermodynamic properties \cite{ref41,ref42,ref43}. For 1D boundary-driven open quantum systems, the different temperature or chemical potential between the two edge baths will cause a current flow from one bath to the other, mediated by the quantum system which can reach  a non-equilibrium steady-state (NESS). The study of the transport properties of such systems, focusing on ways of  controlling the associated current flows efficiently, seems therefore promising owing to potential applications in solid-state devices such as diodes or transistors \cite{ref44}. On the other hand, the dynamical and quantum state transfer properties of such systems \cite{ref45} are also essential from a practical perspective for the implementation of quantum information tasks.

Among the tools available for the study of open quantum systems, tensor network numerical methods based on matrix product states have proven quite efficient for the simulation of their dynamical behaviour \cite{ref46,ref47}. The method of matrix product states can in many cases yield analytical expressions for the expectation values of observables in the NESS of open quantum systems boundary-driven by Lindblad operators for Markovian reservoirs \cite{ref48,ref49,ref50,ref51}. Perturbative approaches \cite{ref52} as well as approaches based on quantum trajectories using numerical Monte Carlo techniques \cite{ref53} have also been explored in great detail.

In this paper, we extend the study of the dynamics of an arbitrary length XX chain boundary-driven by two non-Markovian reservoirs. Our theory is cast in terms of the Schr\"{o}dinger's equations of motion within the single-excitation subspace, which can be solved recursively in Laplace space, yielding closed form solutions for the transformed amplitudes for a broad class of spectral densities of the boundary reservoirs, as well as arbitrary initial conditions. Results on the quantum state transfer properties of the system for various parameters of the two reservoirs are also presented. Our work extends the analytical techniques used for dealing with boundary-driven open quantum systems, by providing a systematic way of calculating the whole wavefunction of the spin chain and therefore enabling the study of numerous properties of the system at hand. 

In much of the extensive literature with methods for the handling of non-Markovian reservoirs, the Lorentzian or Ohmic spectral densities are used for illustration. Yet, the analytical forms of those spectral densities are fairly different. In order to explore the impact of those forms on the behaviour of realistic physical systems, we have in addition performed comparative calculations for those two reservoirs. Our results, discussed in section III seem to reveal what one might call differences in the Markovianity of the two reservoirs.

The article is organized as follows: In section II we provide the theoretical description of our problem in terms of the time-dependent Schr\"{o}dinger equation (TDSE) and obtain closed form solutions for the transformed amplitudes for arbitrary number of sites, arbitrary spectral densities of the boundary reservoirs and arbitrary initial conditions within the single-excitation subspace. In section III we illustrate the potential of our approach through the application to two types of non-Markovian boundary reservoirs, namely Lorentzian and Ohmic, by studying the dynamical or state transfer properties of the system, while in section IV we provide some concluding remarks as well as future directions related to our work.

\section{Theory}
Our system consists of an $N$ qubit (spin) chain interacting with two environments $E_1$ and $E_2$ through its first and last qubit, with coupling strengths $g_1$ and $g_2$, respectively. The interaction between each pair of neighboring qubits in the chain is denoted by $\mathcal{J}$. A schematic presentation of our system is depicted in Fig. 1. The Hamiltonian of our system $\mathcal{H} = \mathcal{H}_S + \mathcal{H}_E + \mathcal{H}_I $, consists of three parts; namely, the Hamiltonian of the XX chain $\mathcal{H}_S$, the Hamiltonian of the two environments $\mathcal{H}_E$ and the chain-environments interaction Hamiltonian $\mathcal{H}_I$, given by the relations $(\hbar = 1)$:

\begin{subequations}
\beq
\begin{split}
\mathcal{H}_S   =  & \omega_e \sum_{i=1}^N \ket{e}_i \prescript{}{i}{\bra{e}}  + \omega_g \sum_{i=1}^N \ket{g}_i \prescript{}{i}{\bra{g}}  \\ 
& + \sum_{i=1}^{N-1} \frac{\mathcal{J}}{2} \left( \sigma_i^{+} \sigma_{i+1}^{-} + \sigma_i^{-} \sigma_{i+1}^{+} \right)
\end{split}
\eeq

\beq
\mathcal{H}_E = \sum_\lambda \omega_{\lambda}^{1} {a^{E_1}_{\lambda}}^{\dagger} a_{\lambda}^{E_1} + \sum_{\lambda} \omega_{\lambda}^{2} {a_{\lambda}^{E_2}}^{\dagger} a_{\lambda}^{E_2}
\eeq

\beq
\begin{split}
\mathcal{H}_I = & \sum_\lambda g_1 ( \omega_{\lambda}^{1} ) \left( a_{\lambda}^{E_1} \sigma_1^{+} +{a_{\lambda}^{E_1}}^{\dagger} \sigma_1^{-} \right) \\
& + \sum_{\lambda} g_2 ( \omega_{\lambda}^{2} ) \left( a_{\lambda}^{E_2} \sigma_N^{+} + {a_{\lambda}^{E_2}}^{\dagger} \sigma_N^{-} \right)
\end{split}
\eeq
\end{subequations}
where $\omega_g$ and $\omega_e$ are, respectively, the energies of the ground and excited state of each spin (all spins are assumed to be identical), $\omega_{\lambda}^i$, $i=1,2$ is the energy of the $\lambda$-mode photon of each environment, $a_{\lambda}^{E_j}$  and ${a^{E_j}_{\lambda}}^{\dagger}$,  $j=1,2$, are the annihilation and creation operators of each environment, respectively, and $\sigma_i^{+}= \ket{e}_i \prescript{}{i}{\bra{g}}$ and $\sigma_i^{-}= \ket{g}_i \prescript{}{i}{\bra{e}}$, $i=1, \dots, N$ are the qubit raising and lowering operators, respectively.

The wavefunction of the compound system (spin chain + environments) in the single-excitation space is expressed as:

\beq
\begin{split}
\ket{\Psi (t)}   = & \sum_{i=1}^N c_i(t) \ket{\psi_i}  + \sum_{\lambda} c_{\lambda}^{E_1}(t) \ket{\psi_{\lambda}^{E_1}} \\
& + \sum_{\lambda} c_{\lambda}^{E_2}(t) \ket{\psi_{\lambda}^{E_2}}
\end{split}
\eeq
where,

\begin{subequations}
\beq
\ket{\psi_i} = \ket{g}_1 \ket{g}_2 \dots \ket{g}_{i-1} \ket{e}_i \ket{g}_{i+1} \dots \ket{g}_N \ket{0}_{E_1} \ket{0}_{E_2}
\eeq

\beq
\ket{\psi_{\lambda}^{1}}= \ket{g}_1 \ket{g}_2 \dots \ket{g}_N \ket{0 0 \dots 0 1_{\lambda} 0 \dots 0 0}_{E_1} \ket{0}_{E_2}
\eeq

\beq
\ket{\psi_{\lambda}^{2}}= \ket{g}_1 \ket{g}_2 \dots \ket{g}_N \ket{0}_{E_1} \ket{0 0 \dots 0 1_{\lambda} 0 \dots 0 0}_{E_2} 
\eeq
\end{subequations}

\begin{figure}[t] 
	\centering
	\includegraphics[width=8.65cm]{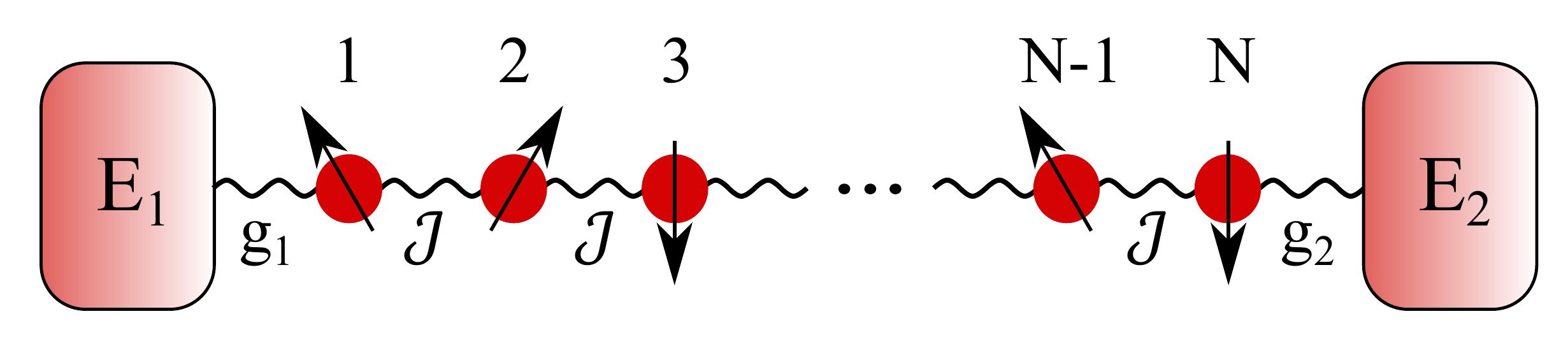}
		\caption[Fig.1]{Schematic presentation of the system at study. A Heisenberg XX spin chain of arbitrary length is coupled to non-Markovian reservoirs at its boundaries.}
\end{figure}
 
 For our derivation, it is useful to adopt the transformations $c_i (t) = e^{-i \left[ \omega_e + (N-1) \omega_g \right] t} \Tilde{c}_i (t)$ , $i = 1, \dots N $ and $ c_{\lambda}^{E_j}(t) = e^ {-i \left( N\omega_g + \omega_{\lambda}^j \right)t} \Tilde{c}_{\lambda}^{E_j}(t)$, $j=1,2$ for the qubits' and environments' amplitudes. The equations of motion of the transformed amplitudes, resulting from the TDSE, are:

\begin{subequations}
\beq
i \frac{ d \Tilde{c}_1 (t)}{dt} = \frac{\mathcal{J}}{2} \Tilde{c}_2 (t) + \sum_{\lambda} g_1(\omega_{\lambda}^{1}) e^{-i \Delta_{\lambda}^{1} t}  \Tilde{c}_{\lambda}^{E_1}(t)
 \label{Amplitude1}
\eeq

\beq
i \frac{ d \Tilde{c}_i (t)}{dt} = \frac{\mathcal{J}}{2} \left[ \Tilde{c}_{i-1} (t) + \Tilde{c}_{i+1} (t) \right], \quad i = 2, \dots, N-1
 \label{Amplitudej}
\eeq

\beq
i \frac{ d \Tilde{c}_N (t)}{dt} = \frac{\mathcal{J}}{2} \Tilde{c}_{N-1} (t) + \sum_{\lambda} g_2(\omega_{\lambda}^{2}) e^{-i \Delta_{\lambda}^{2} t}  \Tilde{c}_{\lambda}^{E_2}(t)
 \label{AmplitudeN}
\eeq

\beq
i \frac{ d \Tilde{c}_{\lambda}^{E_1}(t)}{dt} = g_1(\omega_{\lambda}^{1}) e^{i \Delta_{\lambda}^{1} t} \Tilde{c}_1 (t)
\label{Env1}
\eeq

\beq
i \frac{ d \Tilde{c}_{\lambda}^{E_2}(t)}{dt} = g_2(\omega_{\lambda}^{2}) e^{i \Delta_{\lambda}^{2} t} \Tilde{c}_2 (t)
\label{Env2}
\eeq
\end{subequations}
where $\Delta_{\lambda}^j \equiv \omega_{\lambda}^j - (\omega_e - \omega_g) \equiv \omega_{\lambda}^j - \omega_{eg}$, $j=1,2$.

Formal integration of Eqns. (\ref{Env1}) and (\ref{Env2}), under the assumption that $\Tilde{c}_{\lambda}^{E_1}(0)=\Tilde{c}_{\lambda}^{E_2}(0)=0$,  and substitution back to Eqns. (\ref{Amplitude1}) and (\ref{AmplitudeN}), yields:

\begin{subequations}
\beq
\begin{split}
\frac{ d \Tilde{c}_1 (t)}{d t}  = & -  i   \frac{\mathcal{J}}{2} \Tilde{c}_2 (t) \\
& -  \int_0^t \sum_{\lambda} e^{- i \Delta_{\lambda}^1 \left(  t - t' \right)} \left[  g_1(\omega_{\lambda}^1) \right]^2   \Tilde{c}_1 (t')  dt'
\end{split}
\eeq

\beq
\begin{split}
\frac{ d \Tilde{c}_N (t)}{d t} = & -i \frac{\mathcal{J}}{2} \Tilde{c}_{N-1} (t) \\
& - \int_0^t \sum_{\lambda} e^{- i \Delta_{\lambda}^2 \left(  t - t' \right)} \left[  g_2(\omega_{\lambda}^2) \right]^2   \Tilde{c}_N (t')  dt'
\end{split}
\eeq

\end{subequations}

The summation over all possible modes of each environment can be replaced by an integration that requires the specification of the environment's spectral density $J_j(\omega_{\lambda}^j)$, according to $\sum_{\lambda} \left[  g_j(\omega_{\lambda}^j) \right]^2 \rightarrow \int d \omega_{\lambda}^j J_j (\omega_{\lambda}^j)$, $j=1,2$. In what follows, we keep the spectral density of both environments arbitrary. The resulting set of differential equations then is:  

\begin{subequations}
\beq
\frac{ d \Tilde{c}_1 (t)}{d t}  =  -i  \frac{\mathcal{J}}{2} \Tilde{c}_2 (t) -  \int_0^t R_1 (t-t') \Tilde{c}_1 (t') dt'
\eeq

\beq
 \frac{ d \Tilde{c}_i (t)}{dt} = -i \frac{\mathcal{J}}{2} \left[ \Tilde{c}_{i-1} (t) + \Tilde{c}_{i+1} (t) \right], \quad i= 2, \dots, N-1
\eeq

\beq
\frac{ d \Tilde{c}_N (t)}{d t}  =  -i  \frac{\mathcal{J}}{2} \Tilde{c}_{N-1} (t) -  \int_0^t R_2 (t-t') \Tilde{c}_N (t') dt'
\eeq
 \label{DiffEqns}
\end{subequations}
where we have introduced the definition

\beq
R_j (t) \equiv \int_0^{\infty} J_j (\omega_{\lambda}^j) e^{- i \Delta_{\lambda}^j   t } d\omega_{\lambda}^j, \quad j=1,2
\label{R_j}
\eeq

Taking now the Laplace transform of the above set of differential equations, and using the property of the convolution transform, $\mathcal{L} \left[ \int_0^t f(t') g(t-t') dt'  \right] = F(s) G(s) $, where $F(s)$ and $G(s)$ are the Laplace transforms of the functions $f(t)$ and $g(t)$, respectively, we obtain:

\begin{subequations}
\beq
s F_1(s) = c_1(0) - i\frac{\mathcal{J}}{2} F_2 (s) - B_1 (s) F_1 (s)
\label{LaplaceEq1}
\eeq

\beq
s F_i(s) = c_i(0) - i\frac{\mathcal{J}}{2} \left[ F_{i-1} (s) + F_{i+1} (s) \right], \quad i = 2, \dots, N-1
\label{LaplaceEqj}
\eeq

\beq
s F_N(s) = c_N(0) - i\frac{\mathcal{J}}{2} F_{N-1} (s) - B_2 (s) F_N (s)
\label{LaplaceEqN}
\eeq
\label{LaplaceEqnsAll}
\end{subequations}
where we have used the fact that $\Tilde{c}_i (0) = c_i (0) $,  since $c_i (t) = e^{-i \left[ \omega_e + (N-1) \omega_g \right] t} \Tilde{c}_i (t)$. In the above system of algebraic equations, $F_i (s)$ is the Laplace transform of $\Tilde{c}_i (t)$, while $B_j(s)$ is the Laplace transform of the function $R_j (t)$, $j=1,2$.

As detailed in the Appendix, the set of Eqns. (\ref{LaplaceEqnsAll}), can be organized in such a way that each $F_i(s)$, $i= 2, \dots, N$ is expressed in terms of $F_1(s)$ as:

\beq
\begin{split}
F_i (s) = A_i(s) F_1 (s) + A_{i-1}(s) (ik) B_1 (s) F_1 (s) \\
- (ik) \sum_{n=1}^{i-1} A_{i-n}(s) c_n (0), \quad i = 2, \dots, N
\label{F_i-F_1}
\end{split}
\eeq
where $k \equiv \frac{2}{\mathcal{J}}$ and $A_i(s)$ obeys the relation:
\beq
A_{m+2}(s) = (iks) A_{m+1}(s) - A_m(s), \quad m = 1, \dots, N
\eeq
with $A_1 (s) = 1$ and $A_2 (s) = iks$. Given these conditions, one can easily verify that the $m^{th}$ term of this sequence is given by:

\begin{widetext}
\beq
A_m(s) = \frac{\left[ (iks) +  i \sqrt{k^2 s^2 +4} \right]^m - \left[ (iks) - i \sqrt{k^2 s^2 +4} \right]^m}{2^m i \sqrt{k^2 s^2 +4}}, \\
\quad m = 1, \dots, N
\label{A_m}
\eeq

Taking equation (\ref{F_i-F_1}) for $i=N-1$ and $i=N$, we obtain:
\beq
F_{N-1} (s) = A_{N-1}(s) F_1 (s) + A_{N-2}(s) (ik) B_1 (s) F_1 (s) - (ik) \sum_{m=1}^{N-2} A_{N-1-m}(s) c_m (0)
\label{F_{N-1}}
\eeq
\beq
F_{N} (s) = A_{N}(s) F_1 (s) + A_{N-1}(s) (ik) B_1 (s) F_1 (s) - (ik) \sum_{m=1}^{N-1} A_{N-m}(s) c_m (0)
\label{F_{N}}
\eeq
Substituting Eqns. (\ref{F_{N-1}}) and (\ref{F_{N}}) back into Eqn. (\ref{LaplaceEqN}) and using the relation $\sum_{m=1}^{N-1} A_{N-m}(s) c_m (0) = \sum_{m=1}^{N-2} A_{N-m}(s) c_m (0) + A_1 (s) c_{N-1} (0)$, we can, after some straightforward algebraic manipulations, find that $F_1(s)$ is given by:
\beq
F_1(s) = \frac{ik c_N (0) - k^2 \left[s + B_2 (s) \right] A_1(s) c_{N-1} (0)  + (ik) \sum_{m=1}^{N-2} \Big\{ ik  \left[ s + B_2 (s) \right] A_{N-m}(s) - A_{N-1-m}(s) \Big\} c_m (0)}{ik \left[ s+ B_2 (s) \right] A_N (s) - \Big\{1+k^2 B_1(s) \left[ s+ B_2(s) \right] \Big\} A_{N-1} (s) - ik B_1 (s) A_{N-2} (s)}
\label{F1}
\eeq
\end{widetext}

The inverse Laplace transform of Eqn. (\ref{F1}) provides the time evolution of the amplitude $\Tilde{c}_1 (t)$. The rest of the amplitudes are given by the inverse Laplace transforms of $F_{i} (s)$, $i= 2, \dots, N$ which are recursively related to $F_1(s)$ through Eqn. (\ref{F_i-F_1}).

The analytical method we have developed above allows us to obtain the time evolution of the amplitude of any spin, for an arbitrary number of sites as well as for arbitrary initial conditions, within the single-excitation subspace, by simply calculating the inversion integral of the corresponding Laplace transform. Through our method, the number of sites $N$ has become an arbitrary and easily modifiable parameter entering expression (\ref{F1}). The inversion integral can be readily calculated numerically (or even analytically in some cases), after specification of the spectral density of each environment, necessary for the calculation of the Laplace transforms $B_j(s)$ of the functions $R_j(t)$.

\section{Results \& Discussion}
\subsection{Lorentzian spectral density }

As a first example, we consider the case of an XX chain boundary-driven by two non-Markovian reservoirs with Lorentzian spectral densities, given by:

\beq
J_j (\omega_{\lambda}^j) = \frac{g_j^2}{ \pi} \frac{\frac{\gamma_j}{2}}{\left( \omega_{\lambda}^j-\omega_c^j \right)^2 + (\frac{\gamma_j}{2})^2}, \quad j=1,2
\label{DOS}
\eeq
where $g_j$ are the coupling strengths between the edge spins and the boundary reservoirs in units of frequency, while $\gamma_j$ and $\omega_c^j$, $j=1,2$ are the widths and the peak frequencies of each distribution, respectively.

For analytical simplification of Eqn. (\ref{R_j}), we extend the lower limit of the integration over frequency from 0 to $- \infty$. Note that such extension is not in general valid for any DOS; it is however well justified in the case of a Lorentzian spectral density with positive peak frequency and width such that the distribution has practically negligible extension to negative frequencies. The necessary condition for this is $\gamma_j$ $<<$ $ \omega_c^j$. In this case the frequency integrals can be calculated analytically, yielding:

\beq
\begin{split}
R_j (t) & =  \frac{g_j^2}{\pi} \frac{\gamma_j}{2} \int_{-\infty}^{+\infty} \frac{e^{- i \left( \omega_{\lambda}^j - \omega_{eg} \right)   t  }}{\left( \omega_{\lambda}^j -\omega_c^j \right)^2 + \left( \frac{\gamma_j}{2} \right)^2}  d \omega_{\lambda}^j \\
& = g_j^2 e^{- i \Delta_c^j   t  } e^{-\frac{\gamma_j}{2}   t }, \quad j=1,2
\label{Integration}
\end{split}
\eeq
where $\Delta_c^j \equiv \omega_c^j - \omega_{eg}$, $j=1,2$. The Laplace transforms of $R_j(t)$ are then given by the expressions:

\beq
B_j (s) = \frac{g_j^2}{s+ \frac{\gamma_j}{2} + i \Delta_c^j}, \quad j=1,2
\eeq
Using these expressions of $B_j(s)$, $j=1,2$, we can easily obtain the time evolution of the amplitudes $\Tilde{c}_i(t)$, $i, \dots, N$, using Eqns. (\ref{F_i-F_1}), (\ref{A_m}) and (\ref{F1}). Note that in all of our calculations, we adopt the value $\mathcal{J}=1$ for the strength of the interaction between neighboring qubits, as a reference value for  the comparison with the other couplings entering the formalism.

\begin{figure}[t] 
	\centering
	\includegraphics[width=8.65cm]{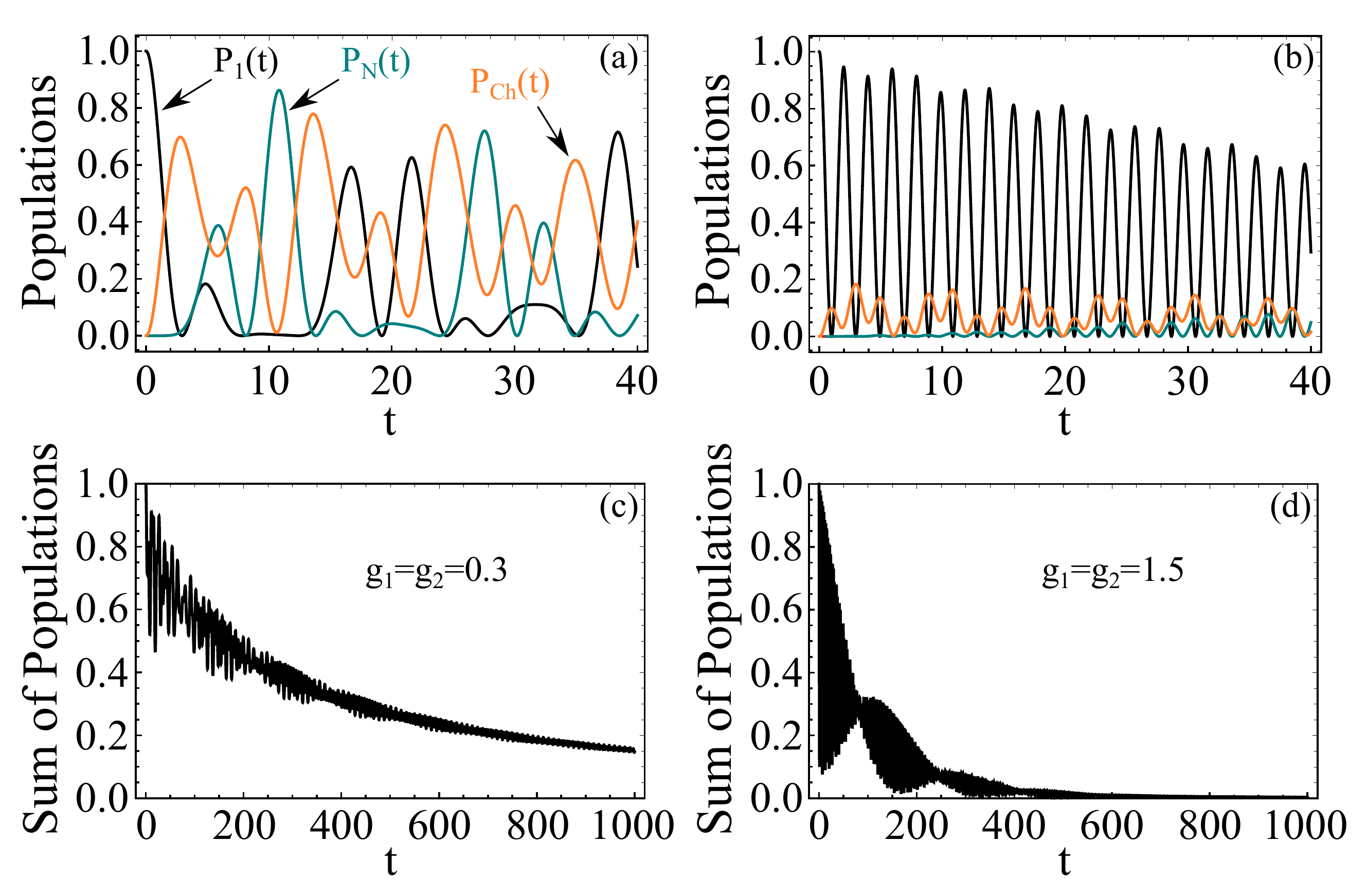}
		\caption[Fig.2]{Dynamics of the populations of the first qubit (black line), $N^{th}$ qubit (teal line) and channel qubits (orange line) for $c_1(0)=1$, $N=5$, $\mathcal{J} = 1$, $\gamma_1=\gamma_2=0.02$, $\Delta_c^1=\Delta_c^2$=0 and (a) $g_1=g_2=0.3$, (b) $g_1=g_2=1.5$. In panels (c) and (d) we show the long-time dynamics of the sum of all qubit populations in the spin chain for $g_1=g_2=0.3$ and $g_1=g_2=1.5$, respectively. }
\end{figure}

\begin{figure*}[t] 
	\centering
	\includegraphics[width=17.3cm]{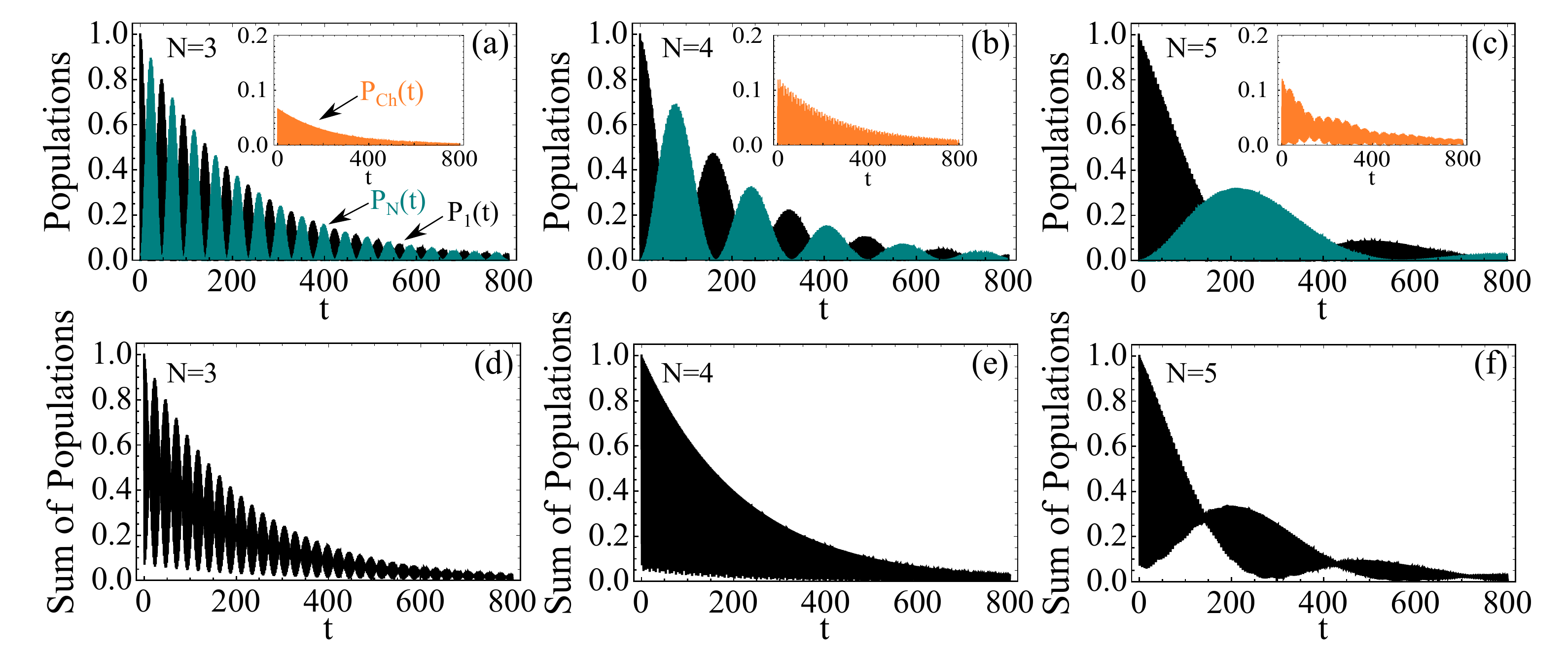}
		\caption[Fig.3]{Long-time dynamics of the populations $P_1(t)$, $P_N(t)$, $P_{Ch}(t)$ (inset) as well as their sum, for spin chains of small length and $c_1(0)=1$, $\mathcal{J} = 1$, $\gamma_1=\gamma_2=0.01$, $g_1=g_2=1.8$, $\Delta_c^1=\Delta_c^2$=0. Panels (a) and (d): $N=3$, panels (b) and (e): $N=4$, panels (c) and (f): $N=5$. }
\end{figure*}

In Fig. 2(a) we study the dynamics of the populations of the edge as well as the intermediate spins in the $g_1$, $g_2$ $<$ $\mathcal{J}$ case for $c_1(0)=1$. For the intermediate qubits, we adopt the term "channel qubits" , which is frequently used in quantum information protocols and represents the sum of the populations of all except the 2 edge qubits. The corresponding occupation probability is $P_{Ch}(t)=\sum_{i=2}^{N-1} | c_i(t) |^2 = \sum_{i=2}^{N-1} | \Tilde{c}_i(t) |^2 $. Since the coupling between the qubits is larger than the couplings between the edge spins and the reservoirs, we observe oscillations in the edge and channel qubit populations, associated with the spreading of the initial excitation throughout the whole chain. Eventually, the population will be lost due to the environmental dissipation. This can be seen through the long-time dynamical picture of the sum of all qubit populations, as in Fig. 2(c). On the the other hand, in the case where $g_1$, $g_2$ $>$ $\mathcal{J}$, as examined in Fig. 2(b), since the initial excitation is on the first qubit, the population is essentially "trapped", albeit not permanently, between the first qubit and environment 1, with only a small portion of the population leaking to the channel. In this case the exact value of $g_2$ is of little relevance since the population of the $N^{th}$ site at any time t is practically negligible. As seen in Fig. 2(d), the strong coupling between the first qubit and the reservoir causes the total population to be lost through the reservoir much faster than in the previous case.

Although, as discussed above, the strong coupling between the environment and the edge qubits may result to population trapping (Fig. 2b), if the chain length is small (up to 5 qubits or so), the long-time dynamics of the chain reveal an oscillatory population transfer between the two edges of the chain (Fig. 3) along with the dissipation to the reservoirs. In that case, the population of the $N-2$ channel qubits remains low for any time t (see inset of Figs. 3a, 3b and 3c). This effect ceases to occur for spin chains of longer length since the excitation cannot be easily transferred between the two edges of the chain. As seen in Fig. 3, the frequency of these oscillation is highly sensitive to $N$, decreasing as the latter increases.

\begin{figure}[H] 
	\centering
	\includegraphics[width=7cm]{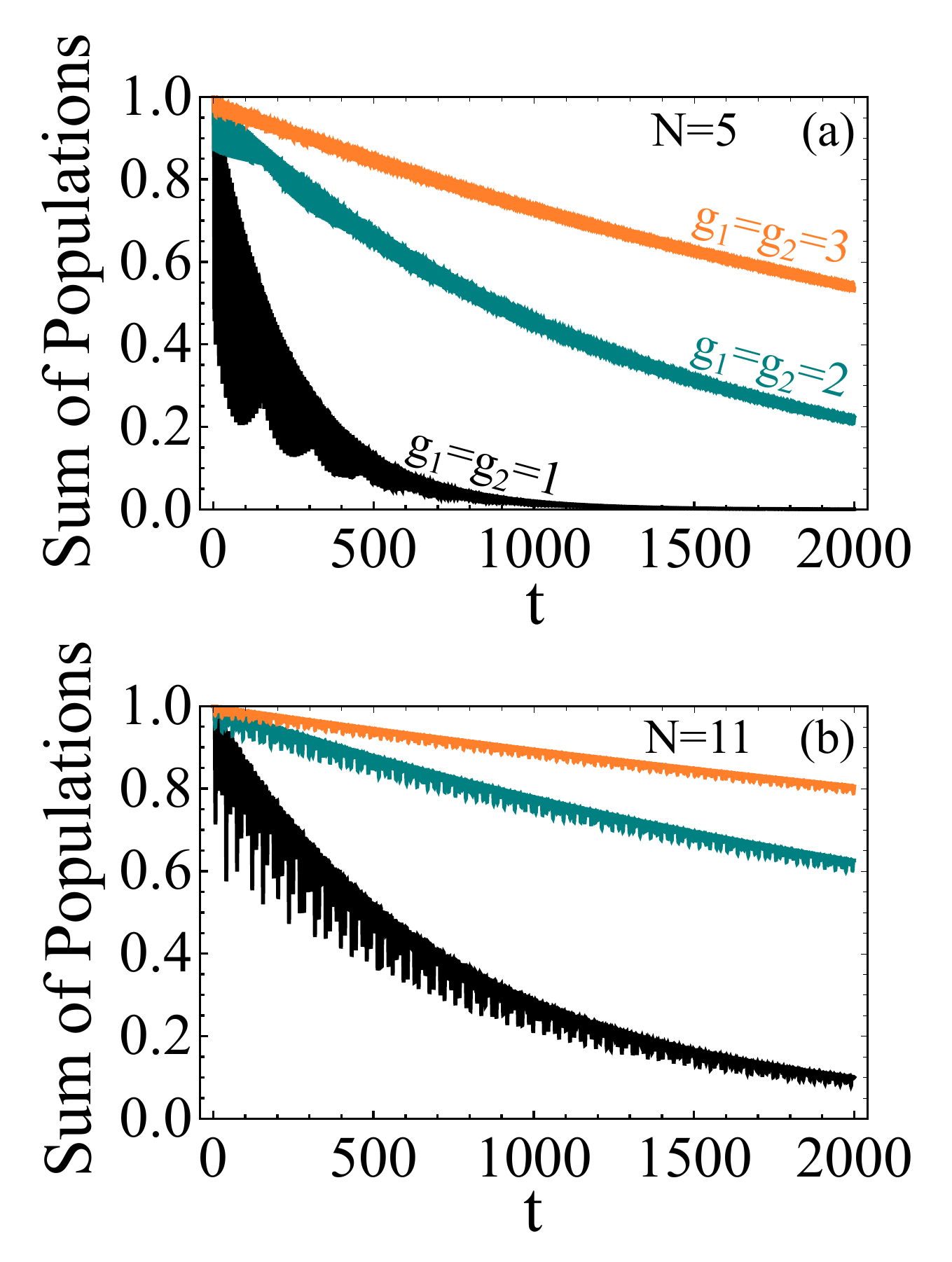}
		\caption[Fig.4]{Long-time dynamics of the sum of all qubit populations in the spin chain with the initial excitation in the central qubit $c_{(N+1)/2}(0)=1$ and $\mathcal{J} = 1$, $\gamma_1=\gamma_2=0.02$, $\Delta_c^1=\Delta_c^2$=0, (a) $N=5$ and (b) $N=11$. Black line: $g_1=g_2=1$, teal line: $g_1=g_2=2$ and orange line: $g_1=g_2=3$.}
\end{figure}

In Fig. 4 we examine the long-time dynamics of the total population for chains with odd number of sites and initial excitation on the centre of the chain, i.e. $c_{(N+1)/2}(0)=1$. In Figs. 2(c) and 2(d) we observed that, if the initial excitation is on the first qubit, the total population of the chain decays much faster for increasing values of $g_1$. This of course is also true for $c_{N}(0)=1$ and increasing values of $g_2$, which is also obvious from symmetry arguments. However, as seen in Fig. 4(a), if the initial excitation is on the centre of the chain, increasing the magnitude of the couplings between the edge spins and the corresponding environment leads to the completely opposite effect, namely a much slower decay of the total population in the long-time dynamics picture. Such an effect occurs only as long as the boundary couplings are larger than the coupling between the qubits $\mathcal{J}$. In the opposite regime, the increase of the boundary couplings (to values still less than $\mathcal{J}$) results to faster decay of the total population to the reservoirs. These results indicate that, for large boundary couplings, the evolution of the edge spins tends to freeze, alluding to a Quantum Zeno effect, leading thus to an effective hindering of the total decay to the reservoirs trough them. This  hindering of the decay becomes even more pronounced for increasing chain lengths (Fig. 4(b)), which physically can be attributed to the increased number of pathways for the evolution of the initial excitation throughout the whole chain and therefore the smaller likelihood for the population to reach  the edge spins and be lost to the reservoirs. 

It may be useful at this point to note that the dissipation of the total population to the reservoirs can also be slowed down by choosing the detunings $\Delta_c^1$ and $\Delta_c^2$ larger than the boundary coupling constants. In that regime of parameters, the dominant modes of the Lorentzian spectral densities of the boundary reservoirs are off-resonant with the qubit frequency and do therefore damp the system less. That might be argued to be obvious. Be that as it may, the statement is valid only in the context of non-Markovian reservoirs. Because the  DOS of a non-Markovian reservoir must at some frequency exhibit a more or less pronounced peak, as it must depart from the slowly varying, smooth DOS which is necessary for the Markovian approximation. It could therefore be viewed as yet another feature of non-Markovian behavior.  

It is also important to note that the effect in which the increased coupling constants between the edge spins and the environments act like a barrier and protect the total population of the chain from decaying to the latter, do not necessary require an initial excitation in the centre of the chain. This is just one among the choices of possible initial conditions where this phenomenon occurs. The general rule is that the initial excitation should be anywhere but the edge spins, in order to avoid the trapping of the population between the edge spins and their corresponding reservoirs, as was the case in Fig. 2(b). For example, the increase of $g_1$ and $g_2$ for a chain system initially prepared in the state $\frac{1}{\sqrt{N-2}}\sum_{i=2}^{N-1}\ket{\psi_i}$, will result to the same phenomenon occurring in Fig. 4.

\begin{figure}[t] 
	\centering
	\includegraphics[width=7.2cm]{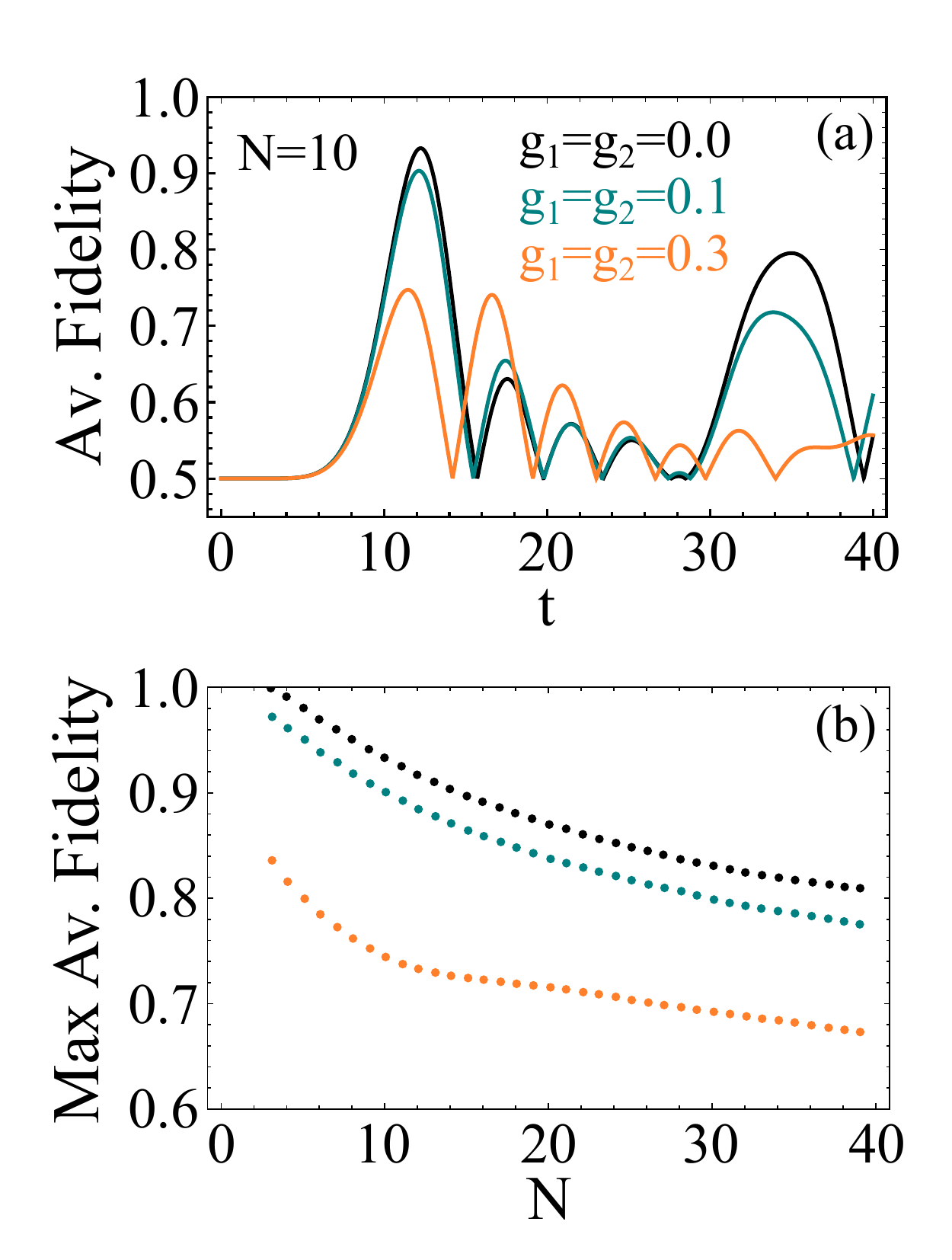}
		\caption[Fig.5]{(a) Dynamics of the average-state fidelity of state transfer between the two edges of the spin chain for $N=10$ and (b) Maximum Average Fidelity as a function of the number of spin sites $N$. In both panels $\mathcal{J}=1$, $\gamma_1=\gamma_2=0.5$, $\Delta_c^1=\Delta_c^2$=0 and black line/dots: $g_1=g_2=0$, teal line/dots: $g_1=g_2=0.1$ and orange line/dots: $g_1=g_2=0.3$.}
\end{figure}

In Fig. 5, we study the state transfer properties between the two edge spins of the chain, in terms of the average-state fidelity \cite{ref54}
\beq
\mathcal{F}(t)= \frac{1}{2} + \frac{|c_N (t)|^2}{6} + \frac{|c_N (t)|}{3} 
\eeq
which involves an average over all possible sender-qubit states. In Fig. 5(a) we show the dynamics of the average-state fidelity for a spin chain with $N=10$ sites, for different values of the boundary couplings $g_1$ and $g_2$. As becomes evident, the first peak of the  average-state fidelity is very sensitive to the boundary couplings, decreasing as their values increase. At the same time subsequent peaks in time tend to increase their maximum value. The position of the first peak, indicating the transfer time is slightly decreased by increasing the boundary couplings, however larger values of the latter lead to non-faithful state transfer in general. In Fig. 5(b) we examine the behaviour of the maximum value of the average-state fidelity as a function of the number of spin sites for chain length up to 40 sites. As $N$ increases, the maximum average-state fidelity tends towards lower values that depend also on the values of the boundary couplings of the edge spins with the reservoirs. Our results qualitatively agree with a similar study by Feng-Hua Ren et al. \cite{ref45}, where they study the state transfer properties of the same system using the dephasing and dissipation models in terms of the quantum state diffusion approach.

\subsection{Ohmic spectral density}

As a second  example,  we  consider  the  case of  boundary reservoirs characterized by Ohmic spectral densities \cite{ref55,ref56}, according to the relation:

\beq
J_j(\omega_{\lambda}^j) = \mathcal{N}_j g_j^2 \omega_c^j \left( \frac{\omega_{\lambda}^j}{\omega_c^j} \right)^{\mathcal{S}_j} \exp \left(- \frac{\omega_{\lambda}^j}{\omega_c^j} \right), \quad j=1,2
\label{OhmicSD}
\eeq
where $g_j$ are qubit-environment coupling constants in units of frequency, $\omega_c^j$ are the so-called Ohmic cut-off frequencies and $\mathcal{S}_j$, $j=1,2$, are the Ohmic parameters, characterizing whether the spectrum of the reservoirs is sub-Ohmic ($\mathcal{S} < 1$), Ohmic ($\mathcal{S}$ = 1) or super-Ohmic ($\mathcal{S} > 1$). $\mathcal{N}_j$ is a normalization constant given by the relation $\mathcal{N}_j= \frac{1}{ \left( {\omega_c^j} \right)^2 \Gamma \left( 1 + \mathcal{S}_j \right)}$, where $\Gamma(z)$ is the gamma function.

Using Eqn. (\ref{OhmicSD}) back to Eqn. (\ref{R_j}), we perform the integration over frequency for arbitrary $\mathcal{S}_j$ and find that $R_j(t)$ is given by the expression:

\beq
\begin{split}
R_j(t) & = \mathcal{N}_j g_j^2 \omega_c^j  \int_0^\infty \left( \frac{\omega_{\lambda}^j}{\omega_c^j} \right)^{\mathcal{S}_j} e^{- \frac{\omega_{\lambda}^j}{\omega_c^j} }  e^{-i \left( \omega_{\lambda}^j- \omega_{eg} \right) t} d \omega_{\lambda}^j \\
& = g_j^2  e^{i \omega_{eg} t} \left( i \omega_c^j t + 1 \right)^{-1-\mathcal{S}_j}, \quad j=1,2
\label{Rj_Ohmic}
\end{split}
\eeq
Note that, unlike much of the literature, in our use of Ohmic spectral densities we found it necessary to include the normalization factor $\mathcal{N}_j$ which renders meaningful the comparison with the Lorentzian spectral density, discussed in subsection IIIC.

The Laplace transform of Eqn. (\ref{Rj_Ohmic}) can also be calculated analytically, yielding:
\beq
\begin{split}
B_j(s) = - g_j^2 \frac{i^{1- \mathcal{S}_j}}{\omega_c^j}   e^{-i K_j(s)}  \left[ K_j(s) \right]^{\mathcal{S}_j} & \Gamma \left(- \mathcal{S}_j, - i K_j(s) \right) \\
& , \quad j=1,2
\label{Bj_Ohmic}
\end{split}
\eeq
where $K_j(s) \equiv \left( s-i \omega_{eg} \right) / \omega_c^j$, $j=1,2$ and $\Gamma(a,z)$ is the incomplete gamma function. In the special case where $\mathcal{S}_j$ are integers, it is useful to use the incomplete gamma function property

\beq
\begin{split}
\Gamma (-n, z) = \frac{1}{n!} \Bigg[ \frac{e^{-z}}{z^n} \sum_{k=0}^{n-1} (-1)^k (n-k-1)! z^k \\
+ (-1)^n \Gamma \left( 0, z \right) \Bigg]
\end{split}
\eeq

which holds for integer n, to express $B_j(s)$ in terms of $\Gamma (0, -iK_j(s))$. Using Eqn. (\ref{Bj_Ohmic}), as in the previous case, we can easily calculate the Laplace inversion numerically and find the time evolution of the amplitudes $\Tilde{c}_i(t)$, $i, \dots, N$.

\begin{figure}[t] 
	\centering
	\includegraphics[width=7cm]{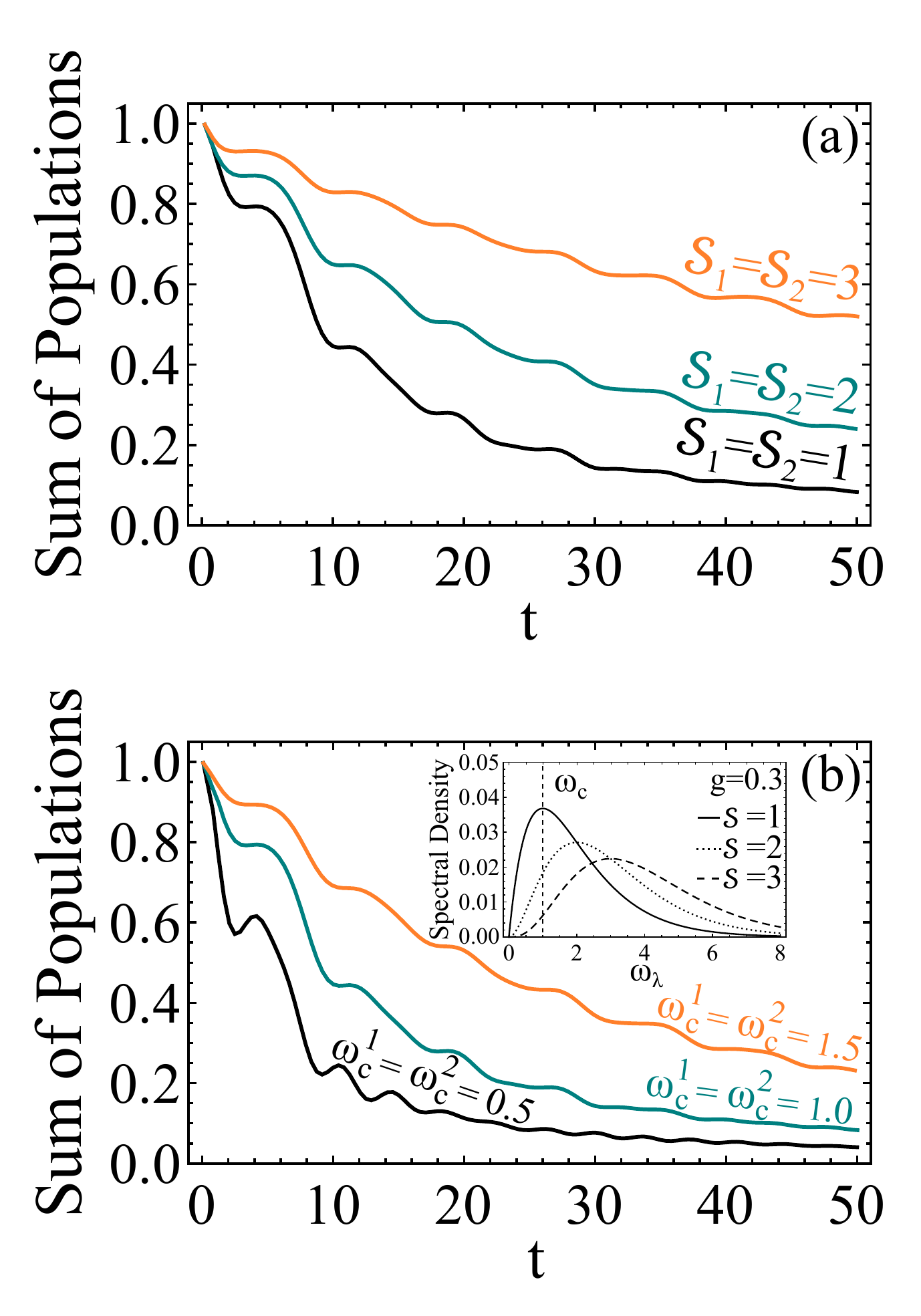}
		\caption[Fig.6]{Dynamics of the sum of all qubit populations in a $N=6$ qubit chain, boundary driven by reservoirs with Ohmic spectral densities. The parameters used in panel (a) are $\omega_c^1=\omega_c^2=1$, black line: $\mathcal{S}_1=\mathcal{S}_2=1$, teal line: $\mathcal{S}_1=\mathcal{S}_2=2$, orange line: $\mathcal{S}_1=\mathcal{S}_2=3$, while in panel (b), $\mathcal{S}_1=\mathcal{S}_2=1$, black line: $\omega_c^1=\omega_c^2=0.5$, teal line: $\omega_c^1=\omega_c^2=1.0$ and orange line: $\omega_c^1=\omega_c^2=1.5$. In both panels, $\mathcal{J}=1$, $g_1=g_2=0.3$ and $\omega_{eg}=1$ and $c_1(0)=1$. The inset in panel (b) shows the Ohmic spectral density as a function of $\omega_{\lambda}$ for various Ohmic parameters $\mathcal{S}$ ($\mathcal{S}=1$: solid line, $\mathcal{S}=2$: dotted line and $\mathcal{S}=3$: dashed line) and $g=0.3$ (coupling strength), $\omega_c=1$.  }
\end{figure}

\begin{figure*}[t] 
	\centering
	\includegraphics[width=14cm]{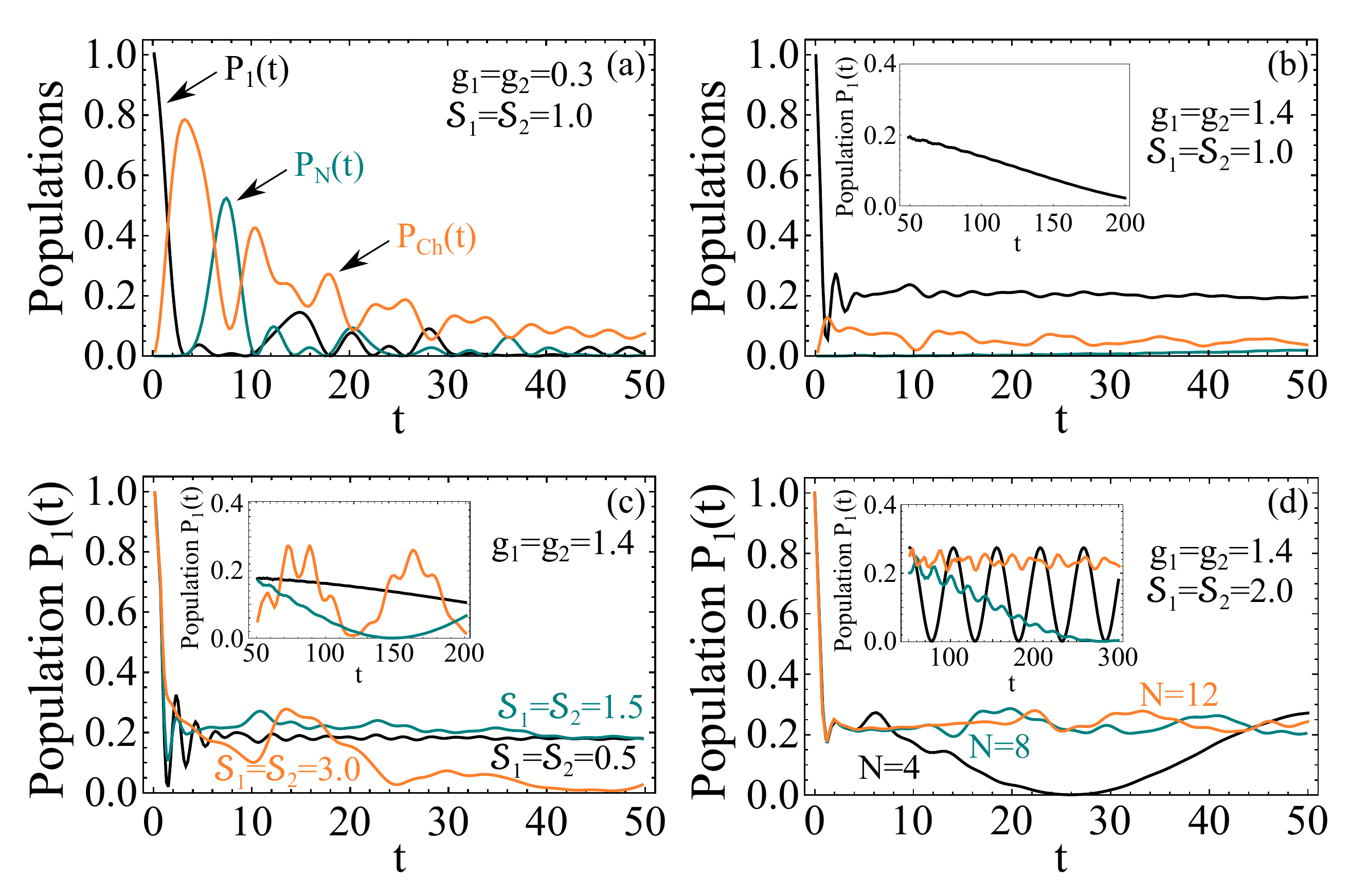}
		\caption[Fig.7]{(a) Dynamics of the populations of the first qubit (black line), $N^{th}$ qubit (teal line) and channel qubits (orange line) for $c_1(0)=1$, $N=6$, $\mathcal{J}=1$, $\omega_{eg}=1$, $\omega_c^1=\omega_c^2=1$, $\mathcal{S}_1=\mathcal{S}_2=1$ and $g_1=g_2=0.3$. (b) Same as in panel (a) but for $g_1=g_2=1.4$. (c) Dynamics of the population $P_1(t)$ for various values of the environmental Ohmic parameters and $c_1(0)=1$, $N=6$, $\mathcal{J}=1$, $\omega_{eg}=1$, $\omega_c^1=\omega_c^2=1$, $g_1=g_2=1.4$. Black line: $\mathcal{S}_1=\mathcal{S}_2=0.5$, teal line: $\mathcal{S}_1=\mathcal{S}_2=1.5$ and orange line: $\mathcal{S}_1=\mathcal{S}_2=3.0$. (d) Dynamics of the population $P_1(t)$ for various values of the number of qubit sites and $c_1(0)=1$, $\mathcal{J}=1$, $\omega_{eg}=1$, $\omega_c^1=\omega_c^2=1$, $\mathcal{S}_1=\mathcal{S}_2=2.0$, $g_1=g_2=1.4$. Black line: $N=4$, teal line: $N=8$, orange line: $N=12$.  In the insets of panels (b), (c) and (d) we show the long-time dynamics of the population $P_1(t)$.}
\end{figure*}

In Fig. 6 we study the dynamics of the sum of all qubit populations in the chain for various parameters of the Ohmic reservoirs. The dynamics of this quantity provide us straightforward information of how fast the initial excitation gets lost in the two environments. As Fig. 6(a) indicates, by increasing the Ohmic parameters $\mathcal{S}_j$, $j=1,2$ of the reservoirs, the sum of populations decays more slowly,  indicating a slower decay of the single excitation to the environments. Physically, this effect is attributed to the fact that by increasing $\mathcal{S}_j$, the spectral density distribution of the environments is mainly peaked in the vicinity of modes whose frequency is larger than the qubit frequency $\omega_{eg}=1$ (see inset of Fig. 6(b)), and therefore by being off-resonant with the qubit frequency, they damp the system less. A similar behaviour is observed in Fig. 6(b), where the dynamics are shown as function of the cut-off frequencies, for Ohmic type of environments ($\mathcal{S}_j$=1). Since for Ohmic distributions the peak of the distribution coincides with the cut-off frequency $\omega_c^j$, $j=1,2$, increase of the cut-off frequencies results to distributions whose modes extend mainly beyond $\omega_{eg}$, resulting to a slower decay to the boundary environments. Note that in general, for arbitrary Ohmic parameters, the Ohmic distributions are always peaked at the frequencies $\omega^j_{\lambda}= \mathcal{S}_j\omega_c^j$, $j=1,2$. The characteristic oscillations observed in both panels of Fig. 7 are indicative of the non-Markovian character of the Ohmic boundary reservoirs, enabling the exchange of population between them and the spin chain within finite times.

In Fig. 7(a) we examine the dynamics of the populations of the edge as well as the channel qubits of the chain for Ohmic boundary reservoirs ($\mathcal{S}_j$=1, $j=1,2$) in the weak boundary couplings regime $g_1, g_2 < \mathcal{J}$. The initial single-excitation is on the first qubit ($c_1(0)=1$). The dynamics indicate oscillations between the populations of the chain damped by the two Ohmic environments. The situation however may change drastically in the strong boundary couplings regime $g_1, g_2 > \mathcal{J}$, as shown in Fig. 7(b). In this case the dynamics of the population $P_1(t)$ indicate a long-time stability behaviour around a finite value for times much larger than any other time scale of the system. Note that further increase of the boundary coupling constants will not result to the increase of this long-time stability value. Still, as seen in Fig. 7(c), this value is highly affected by the Ohmic parameters of the environments (more specifically by the Ohmic parameter of the first environment which is the one that communicates with the first qubit). In general, increase of the Ohmic parameters results to more complex behaviour as indicated by the orange line of Fig. 7(c). In the long-time dynamics picture, the population of the first qubit exhibits an oscillatory behaviour with a frequency that depends on the Ohmic parameter of its boundary reservoir. These non-coherent oscillations become faster for super-Ohmic reservoirs and are essentially an effect associated with the non-Markovianity of the boundary reservoirs that drive the chain, since they indicate population exchanges between the chain and the latter for very long times. Another parameter that affects considerably this oscillatory behaviour is the number of qubit sites. In Fig. 7(d) we plot the dynamics of $P_1(t)$ for a chain of various number of qubit sites $N$, boundary-driven by super-Ohmic environments with $\mathcal{S}_j$=2, $j=1,2$. For a small number of qubit sites, the long-time dynamics in the inset of Fig. 7(d) indicate a long-living coherent oscillatory behaviour (black line) while for chains that consist of larger number of qubits, the population of the first qubit exhibit small non-coherent oscillations around a stabilized value (orange line). Our results are just a first glimpse on the richness of the effects one may expect from the boundary-driving of chains with Ohmic type of reservoirs.

\begin{figure}[t] 
	\centering
	\includegraphics[width=7cm]{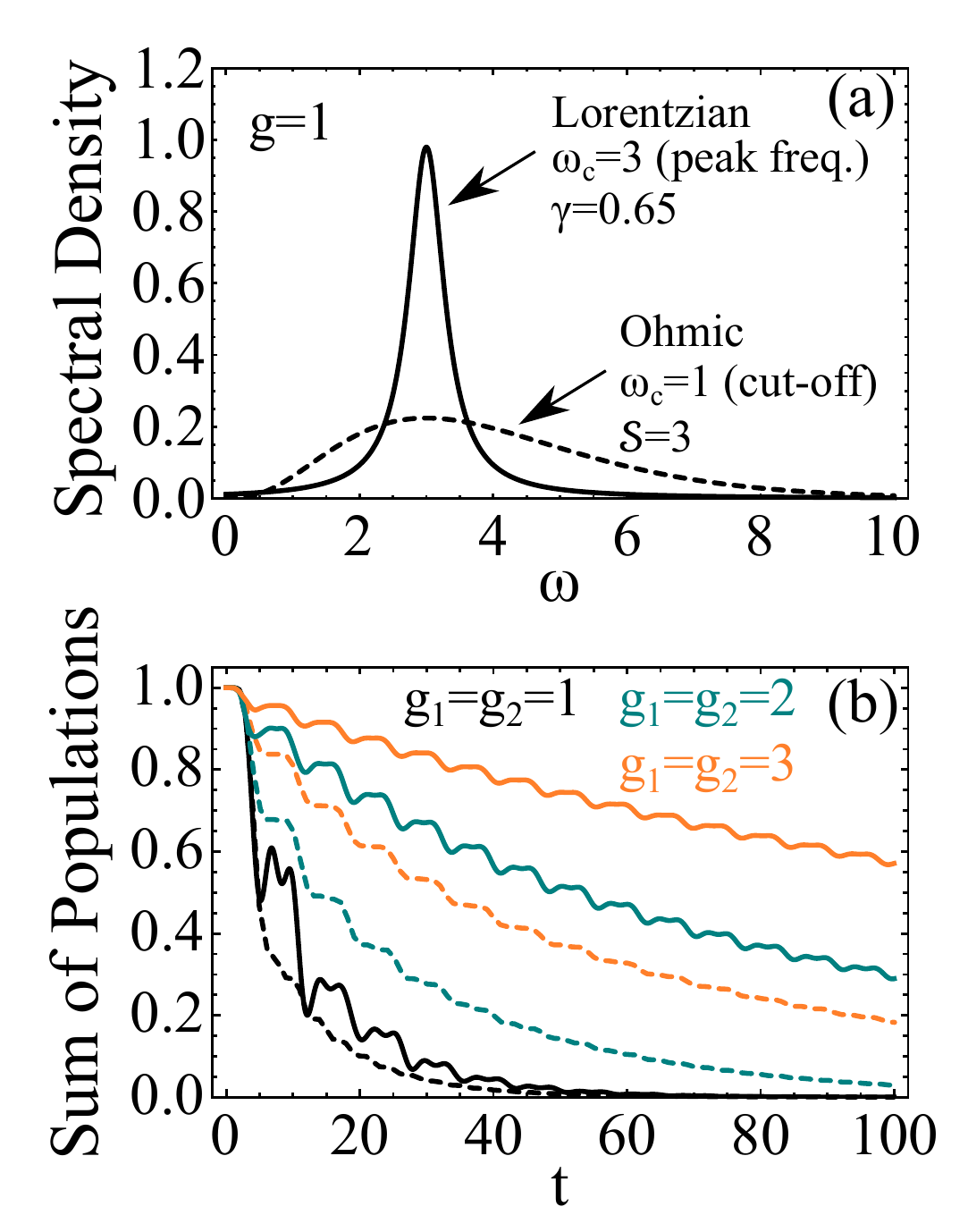}
		\caption[Fig.8]{(a) Spectral density of a Lorentzian distribution  with peak frequency $\omega_c=3$, width $\gamma=0.65$ and $g=1$ (solid line) compared with an Ohmic distribution with Ohmic parameter $\mathcal{S}=3$, cut-off frequency $\omega_c=1$ and $g=1$ (dashed line).  (b) Dynamics of the sum of all qubit populations in a N=7 chain and initial excitation in the center of the chain ($c_{(N+1)/2}(0)=1$), for various coupling strengths between the boundary reservoirs and the edge qubits. Black line: $g_1=g_2=1$, teal line: $g_1=g_2=2$ and orange line: $g_1=g_2=3$. The solid lines corresponds to Lorentzian boundary reservoirs, while the dashed lines correspond to Ohmic. The rest of the parameters (except the coupling strengths) correspond to the parameters of panel (a). The qubit frequency is chosen to be resonant with the peak frequency of the two distributions, namely $\omega_{eg}=3$.}
\end{figure}

\subsection{Comparing Lorentzian to Ohmic spectral densities}

So far we have separately studied the cases of boundary-driven chains by reservoirs with Lorentzian and Ohmic spectral densities. In Fig. 8 we provide a case for comparison between the two. As seen in Fig. 8(a) we have chosen both distributions to be peaked exactly on resonance with the qubit frequency $\omega_{eg}=3$. Although the freedom in the choice of the parameters of each distribution, results to many possible lineshapes of their spectral densities, the general tendency for typical parameters of the two distributions is that Ohmic distributions are usually wider than Lorentzian distributions. In Fig. 8(b) we use the distributions of Fig. 8(a) as spectral densities of the boundary reservoirs and study the dynamics of the sum of all qubit populations in a $N=7$ chain for various values of the boundary couplings $g_j$, $j=1,2$ and initial excitation in the center of the chain. The solid lines correspond to Lorentzian boundary reservoirs while the dashed lines correspond to Ohmic. As also argued in the discussion of Fig. 4, since the initial excitation is not on the edge qubits of the chain and the boundary couplings are larger than the qubit-qubit coupling $\mathcal{J}$, increasing the values of $g_j$, $j=1,2$, results to slower dissipation of the total excitation in the environments. Comparison between the solid (Lorentzian) and dashed (Ohmic) lines, reveals two major differences: First, the dynamics of the total population in the boundary-driven chain by Lorentzian reservoirs is more sensitive to the increase of the boundary couplings compared to the Ohmic-driven chain. Moreover, for the particular combination of parameters chosen, the total excitation in the chain tends to live longer in Lorentzian-driven chains compared to Ohmic-driven ones, which can be attributed to the much broader profile of the Ohmic spectral density compared to the Lorentzian one. Second, it is evident that the dynamics of the sum of all qubit populations exhibit more vivid oscillations in the Lorentzian case compared to the Ohmic one. Such oscillations are indicative of the non-Markovian character of the boundary-reservoirs and therefore the exchange of populations between the chain and the reservoirs within finite times. Even if both types of reservoirs are non-Markovian, the Ohmic spectral density of Fig. 8(a) tends to be substantially flatter than the Lorentzian spectral density, resembling thus a Markovian environment for  which  no oscillations are expected between the chain and its boundaries.

Finally, before concluding, we should note that the dynamics of the average-state fidelity of the state transfer between the edges of a chain boundary-driven by Ohmic reservoirs display qualitatively more or less similar behaviour as the corresponding average-state fidelity dynamics of a boundary-driven chain by Lorentzian reservoirs, as in Fig. 5.

\section{Concluding Remarks}

In this paper, we present a formalism for the systematic way of calculating the wavefunction of an arbitrary length XX spin chain, boundary-driven by non-Markovian environments of arbitrary spectral density. The theory is cast in terms of Schr\"{o}dinger's equations of motion, within the single-excitation subspace, and leads to closed form solutions for the Laplace transforms of the amplitudes for arbitrary initial conditions, using a recursive method. The inversion integrals can be easily calculated numerically (or even analytically in some cases), upon specification of the spectral density of each environment. The calculation of the wavefunction of the chain within the single-excitation subspace enables the study of a number of interesting properties of the system. As an illustration of the potential of the approach, we consider Lorentzian and Ohmic spectral densities for the boundary environments and study aspects of the dynamical, as well as of the state transfer properties of the system.  Our results demonstrate the memory effects of the non-Markovian environments. The concomitant information exchange between the latter and the spin chain, give rise to a plethora of interesting effects such as population trapping, oscillations between the populations of the edge spins or the protection of the spin populations against dissipation for increased values of the boundary couplings.

One aspect of our results that turns out to be revealing and somewhat surprising has to do with the comparative analysis of the dynamics of the chain under two quite different reservoirs; namely Lorentzian and Ohmic. We adopt the general notion that any reservoir, with spectral density not smoothly and slowly varying as a function of energy (frequency), in an extended range around the resonance with the qubit, is non-Markovian. This entails some degree of memory, which in turn implies that the loss into the reservoir would not be monotonic. In much of the work and approaches on non-Markovian reservoirs, it is the Lorentzian that has served as a typical illustration. Physically, that behavior should be expected for a Lorentzian spectral density as it exhibits a well-defined peak. But so does an Ohmic as well as a superohmic spectral density. However, our direct quantitative comparison of the effect of a Lorentzian DOS versus that of a superohmic in section IIIC, reveals effects of an almost qualitative difference between the two; the effect of the superohmic resembles an almost Markovian reservoir, which implies a much lower degree of non-Markovianity. This concrete example complements earlier general formulations of measures of non-Markovian behavior \cite{ref57}.

Our work was motivated by the rapidly growing interest in the field of boundary-driven open quantum systems and in particular the need for the development of the necessary analytical or numerical tools for the study of their properties. The results provide the background for a number of  extensions, such as accounting for different spin chain configurations, as well as the theoretical description of the system, within our formal  development, beyond the single-excitation. It would in addition be interesting to investigate whether the behavior for long times of the edge qubits, which communicate directly with the boundary reservoirs (Fig. 7), could be useful in the survival of long-distance quantum correlations, for chains consisting of large number of qubit sites, boundary-driven by Ohmic reservoirs.

\section*{Acknowledgments}
GM would like to acknowledge the Institute of Electronic Structure and Laser (IESL), FORTH, for financially supporting this work through its Research and Development Program $\Pi \Delta$E00710. We are also thankful to Prof. Martin B. Plenio for his critical reading of the manuscript.

\appendix
\renewcommand{\thesection}{\Alph{section}}
\numberwithin{equation}{section}

\section{}

The system of Eqns. (\ref{LaplaceEqnsAll}) is a set of $N$ algebraic equations that can be solved recursively as follows: First we solve Eqn. (\ref{LaplaceEq1}) for $F_2 (s)$ and Eqn. (\ref{LaplaceEqj}) for $F_{i+1} (s)$ to find
\begin{widetext} 
\begin{subequations}
\beq
F_2(s) = (iks) F_1(s) + (ik) B_1(s) F_1(s) - (ik) c_1 (0)
\eeq

\beq
F_{i+1}(s) = - F_{i-1}(s) + (iks) F_i(s) - (ik) c_j(0), \quad i = 2, \dots, N-1
\eeq
\end{subequations}
where $k \equiv 2/\mathcal{J}$.
Using these two relations, we can express each $F_i(s)$ in terms of $F_1(s)$. The first few terms up to $F_6(s)$ can be organized as follows:

\begin{subequations}
\beq
F_2(s) = (iks) F_1(s) + (ik) B_1(s) F_1(s) - (ik) c_1 (0)
\eeq
\beq
F_3(s) = \left[ (iks)^2 -1 \right] F_1 (s) + (iks) (ik)  B_1(s) F_1(s) - (ik) \left[ (iks) c_1 (0) + c_2 (0) \right]
\eeq
\beq
F_4(s) = \left[ (iks)^3 -2 (iks) \right] F_1 (s) + \left[ (iks)^2 -1 \right](ik)  B_1(s) F_1(s) - (ik) \Big\{ \left[ (iks)^2 -1 \right] c_1 (0) + (iks) c_2 (0) -  c_3 (0) \Big\}
\eeq
\beq
\begin{split}
F_5(s) = & \left[ (iks)^4 - 3 (iks)^2 + 1 \right] F_1 (s) + \left[ (iks)^3 -2 (iks) \right] (ik)  B_1(s) F_1(s) \\
& - (ik) \Big\{ \left[ (iks)^3 -2 (iks) \right] c_1 (0) + \left[ (iks)^2 -1 \right] c_2 (0) + (iks) c_3 (0) + c_4 (0) \Big\}
\end{split}
\eeq
\beq
\begin{split}
F_6(s) = & \left[ (iks)^5 -4 (iks)^3 + 3 (iks) \right] F_1 (s) + \left[ (iks)^4 -3 (iks)^2 +1 \right] B_1(s) F_1(s) \\ 
& - (ik) \Big\{ \left[ (iks)^4 -3 (iks)^2 +1 \right] c_1 (0) + \left[ (iks)^3 -2 (iks) \right] c_2 (0) + \left[ (iks)^2 -1 \right] c_3 (0) + (iks) c_4 (0) + c_5 (0) \Big\}
\end{split}
\eeq
\end{subequations}

Careful inspection of the above system of equations reveals that $F_i (s)$, follows a pattern of the form:

\beq
F_i (s) = A_i(s) F_1 (s) + A_{i-1}(s) (ik) B_1 (s) F_1 (s) - (ik) \sum_{n=1}^{i-1} A_{i-n}(s) c_n (0), \quad i = 2, \dots, N
\eeq
where $A_i(s)$ obeys the relation:
\beq
A_{m+2}(s) = (iks) A_{m+1}(s) - A_m(s), \quad m = 1, \dots, N
\eeq
with $A_1 (s) = 1$ and $A_2 (s) = iks$. The $m^{th}$ term of this sequence is given by:

\beq
A_m(s) = \frac{\left[ (iks) +  i \sqrt{k^2 s^2 +4} \right]^m - \left[ (iks) - i \sqrt{k^2 s^2 +4} \right]^m}{2^m i \sqrt{k^2 s^2 +4}}, \quad m = 1, \dots, N
\eeq
\end{widetext}

\end{document}